%% file: IEEE_paper_main.tex
\documentclass[%
	final,
	journal,
    comsoc,
	letterpaper,
	oneside,
	twocolumn,
	nofonttune,%
]{IEEEtran}%
\input{./organization/preamble.tex}%
\input{./organization/makros.tex}%
\input{./organization/settings.tex}%
\usepackage[letterpaper, left=0.6in, right=0.6in, bottom=1in, top=0.7in]{geometry}
\usepackage{algorithm}
\usepackage{algpseudocode}
\usepackage{newtxmath}
\usepackage{booktabs}
\makeatletter
\let\blx@rerun@biber\relax
\makeatother
				\newcommand{\disablewr}[1]{#1}%
				\newcommand{\newcommanddisw}[3]{\newcommand{#1}[1]{\disablewr{\textcolor{#2}{#3}}}}%
\renewcommand{\disablewr}[1]{}%
\definecolor{todocol}{named}{red}
\newcommanddisw{\todo}{todocol}{ToDo: #1}%
\definecolor{migucol}{named}{purple}%
\newcommanddisw{\migucom}{migucol}{{@}comment: #1}%
\newcommanddisw{\miguhigh}{migucol}{#1}%

\definecolor{darecol}{named}{blue}%
\newcommanddisw{\darecom}{darecol}{{@}comment: #1}%
\newcommanddisw{\darehigh}{darecol}{#1}%

	\newcommand{\TempDisplayPreparation}{\disablewr{%
		\section{Draft-State: Comment Color Code}\noindent%
		\todo{Comments: ToDos}\nl%
		\migucom{To Do and Comments: Michael Gundall}\nl%
		\darecom{To Do and Comments: Daniel Reti}
	}}%
\begin{document}%
%
\title{%
Implementation and Evaluation of the RBIS Protocol in 5G
\thanks{This is a preprint of a work accepted but not yet published at the 2022 IEEE Globecom Workshops (GC Wkshps). Please cite as: M. Gundall, J. Stegmann, C. Huber,  R. Halfmann, and H.D. Schotten: “Implementation and Evaluation of the RBIS Protocol in 5G”. In: 2022 IEEE Globecom Workshops (GC Wkshps), IEEE, 2022.}
}%
%
%
\input{./organization/IEEE_authors-long.tex}%
%
%
%
%
%
%
%
\maketitle
\renewcommand{\figurename}{Fig.}
\renewcommand{\tablename}{Tab.}
%
%
%
\begin{abstract}%
5G that is strongly focused on improvements on QoS and QoS guarantees, which are necessary for industrial deployments, enables novel use cases. Here, nearly each use case requires time synchronization of the involved systems. While PTP in its variations, e.g. IEEE~1588 v2.1 or IEEE~802.1AS, has established as standard for wireline systems, time synchronization of wireless or hybrid systems is still subject to research. 

Thus, the so-called RBIS protocol, which was originally developed and investigated for \mbox{Wi-Fi}, is mapped to 5G. This is possible, because both systems are infrastructure based and a suitable broadcast that fits to the requirements of RBIS protocol can be found in the  control layer of 5G~NR. Even if the $1~\mu$s requirement that is required by some applications is not yet cracked, the accuracy of $1.3~\mu$s and precision of $\leq 4.3~\mu$s for non-invasive extension of existing 5G deployments is highly promising.
\end{abstract}%
%
%
%
%
%
\IEEEpeerreviewmaketitle
%
%
%
%
%
%
%
%

\section{Introduction}%
\label{sec:Introduction}
\gls{5g} 
are the basis for future wireless applications. Thus, \gls{5g} systems recently enter factory halls to firstly solve stringent wireless industrial use cases \cite{gundall20185G}. 
The continuous integration of systems into the wireless landscape as well as the rising demands on \gls{qos} lead to more and more distributed systems. In contrary, dependant on specific applications, a dynamic system composition is feasible. 
Here, a high synchronization quality between the systems is urgently required. While clock synchronization is mostly solved for wireline systems, it is still under research for wireless or hybrid communication systems. As precise clock synchronization is already a challenge for wireless systems, hybrid systems demand the convergence of wired and wireless systems. This requires the integration of an existing time domain, e.g., a \gls{tsn} network. While a transparent tunneling of \gls{ptp} messages via \gls{5g} \glspl{upf} is possible, the performance is highly decreased due to both determinism of the air interface and the \gls{upf} scheduling in the core network.         

Overall, the most demanding requirements regarding synchronicity are local devices that have to cooperate \cite{9613325}, whereas most challenging use cases require a clock synchronicity of $\leq$1$\mu$s \cite{9204594}. Furthermore, an increased synchronization between these devices not only leads to realization of novel use cases or improved product quality, but also to higher possible speed and product processing, e.g. by mobile robots. This leads to both increased efficiency and revenue. 





Consequently, the following contributions can be found in this paper:
\textbf{
\begin{itemize}
    \item Concept for clock synchronization of \gls{5g} \glspl{ue} on the basis of \gls{rbis} protocol. 
    \item Evaluation of the proposed concept performing measurements on a testbed. 
\end{itemize}
}

Accordingly, the paper is structured as follows: 
Sec. \ref{sec:Clock synchronization} describes the principles and the state-of-the-art regarding clock synchronization, while Sec. \ref{sec:Related Work} gives an overview of related work. Furthermore, Sec.~\ref{sec:Mapping of RBIS to 5G} proposes the mapping of the originally for \mbox{Wi-Fi} designed \gls{rbis} protocol to \gls{5g}. 
Furthermore, the testbed is introduced (Sec.~\ref{sec:Testbed}) and the concept is evaluated (Sec.~\ref{sec:Analysis / Evaluation}). Finally, the paper is concluded (Sec. \ref{sec:Conclusion}).

\section{Clock synchronization}%
\label{sec:Clock synchronization}

\begin{figure}[htbp]
\centering
  \includegraphics[width=.7\columnwidth]{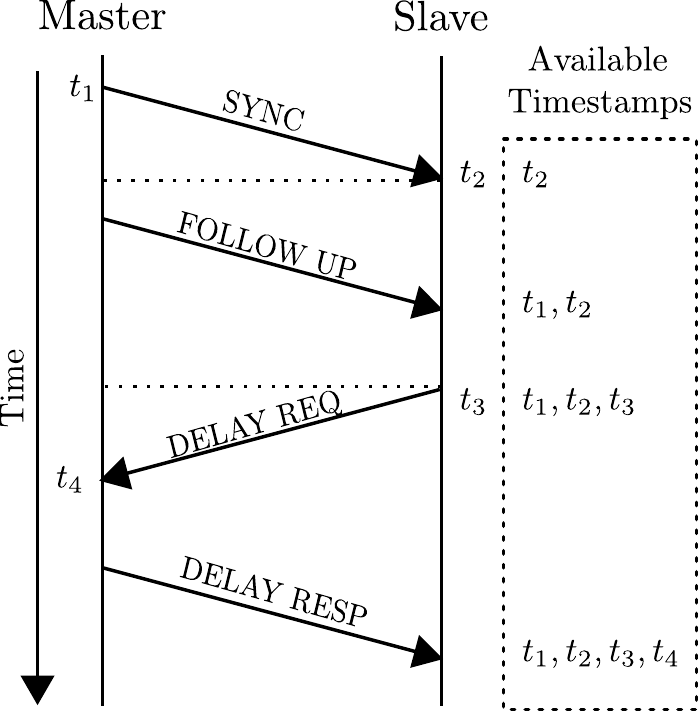} 
  \caption{Message exchange and corresponding timestamps in PTP (IEEE 1588).}
  \label{fig:ptpv2}
\end{figure}

A precise and accurate clock synchronization is the basis for a wide range of industrial use cases and applications. As of now, most industrial communication systems are wireline, due to several advantages, such as determinism, reliability, small and bounded latency, and more. Thus, lots of clock synchronization algorithms were developed specifically for this type of communication, whereby the most common ones can be grouped into one-way and two-way synchronization protocols \cite{9145977}. 
 While one-way synchronization protocols, such as \gls{ftsp}, send only one message from the time master to one or more slaves, the quality is highly dependent on the latency and jitter of the \gls{lan} \cite{maroti2004flooding}. Therefore, two-way synchronization protocols, such as \gls{ntp} were introduced \cite{maroti2004flooding}. 
 Using \gls{ntp}, a single message exchange following the ping-pong scheme and containing several timestamps can be used to reach a \mbox{10-100~ms} accuracy in wireline systems, depending strongly on the number of network devices located in the \gls{lan} \cite{4077944}.  While this accuracy is often sufficient for non-industrial end users, most operating systems of \glspl{pc} are synchronized using this approach. However, for industrial environments, a higher accuracy is required. 
Hence, the \gls{ptp} (IEEE 1588) was introduced \cite{eidson2002ieee}. As shown in Fig.~\ref{fig:ptpv2}, a total of four messages are sent. 
While \textit{SYNC} and \textit{DELAY~REQ} are used as trigger for creating the timestamps, \textit{FOLLOW~UP} and \textit{DELAY~RESP} make the timestamps available for the slave \cite{4077944}.  During the complete process four different timestamps are produced. With the help of these timestamps, not only the offset between master and slave clock can be calculated (see Eq.~\ref{eq:offsetptp}), but also the delay of the packet transmission (see Eq.~\ref{eq:delay}). 

\begin{equation}
Delay = \frac{(t_2-t_1)+(t_4-t_3)}{2} \label{eq:delay}
\end{equation}

\begin{equation}
Offset = \frac{(t_2-t_1)-(t_4-t_3)}{2} \label{eq:offsetptp}
\end{equation}

Since the delay measurement and its correction highly improves the clock synchronization, it is widely used whenever a precise clock is required. Due to the fact that \gls{ptp} is based on delay measurements, the best performance can be reached in local wireline 
networks, where the path of the packet and its delay is nearly static.
Thus, in a \gls{lan} easily microsecond or even sub-microsecond accuracy could be reached, whereas the actual performance depends on the capabilities of the devices, i.e., master and slave providing hardware timestamping or the availability of network switches with \ac{ptp} support. Hence, the recently developed \gls{tsn} technology 
specifies a standard for time synchronization using \gls{ptp} as basis \cite{8412459}. This so-called IEEE~802.1AS standard extends \gls{ptp} by a novel configuration. In addition, hardware timestamping capabilities and protocol stack message exchange directly at the MAC layer are required to comply with this standard. Using these features, nanosecond accuracy is possible. 
Since some use cases even require sub-nanosecond accuracy the IEEE~1588~v2.1 profile supports this accuracy, but requires sophisticated devices and extensions \cite{6070148}.

\begin{figure}[htbp]
\centering
\includegraphics[width=.7\columnwidth]{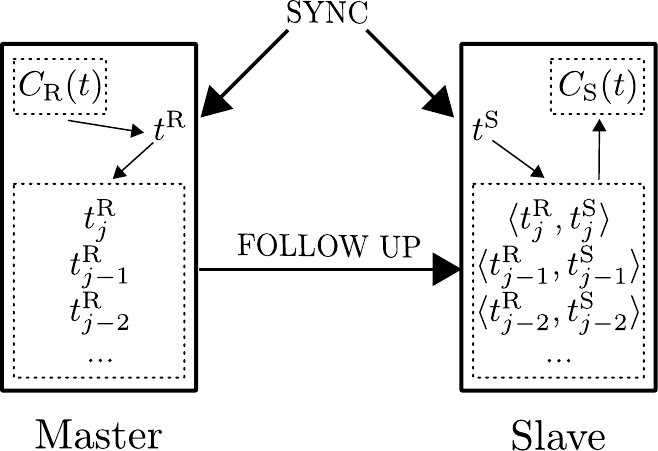} 
\caption{\gls{rbis} protocol (refined from \cite{cena2018neural}).}
\label{fig:rbis}
\end{figure}

While \gls{ptp} and its deviations are well suited for clock synchronization in wireline systems, clock synchronization in wireless systems requires different approaches, since the introduced delay measurement is not as accurate as for wireline systems. This is because the propagation paths of the signals sent from mobile devices change frequently during operation. Thus, different concepts are required for the clock synchronization of devices using wireless communication links.
Here, the so-called \gls{rbs} protocol, which belongs to the group of one-way synchronization protocols, was developed \cite{ganeriwal2003timing}. In the \gls{rbs} protocol one node acts as master and sends a cyclic broadcast including the reference time to all slaves that have to  be synchronized. In this way it makes use of the properties of the shared communication channel because it is assumed that the broadcast arrives at each of the slaves at nearly the same time. This assumption is possible, because the range of a \mbox{Wi-Fi} deployment is typically limited. Since the purpose of the \gls{rbs} protocol is to synchronize infrastructure-free networks, but many novel wireless communication systems such as \mbox{Wi-Fi} or \gls{5g} are infrastructure-based, a separate version, the \gls{rbis} protocol, was introduced  \cite{7018946}, as shown in Fig.~\ref{fig:rbis}. 

In this case the infrastructure sends the cyclic synchronization broadcast and the connected nodes are synchronized, whereas the infrastructure cannot be synchronized using this approach. When the devices get the broadcast each device tags the incoming synchronization message to their recent timestamp that is device specific. Afterwards, the device that acts as master can distribute its reference timestamps $t_j^R$ to all slaves as \textit{FOLLOW~UP} message. Then, all slave devices have both, their own and reference timestamps  $\langle t_j^S, t_j^R \rangle$ for a specific synchronization message. Thus, the offset between master and slave clocks $C_R(t)$ and $C_S(t)$ can be calculated.

\section{Related Work}%
\label{sec:Related Work}
The benefits of \gls{rbis} protocol already motivated several investigations. Thus, \cite{7018946} 
evaluated the concept based on \mbox{\mbox{Wi-Fi}}. 
Since the \gls{rbis} protocol requires cyclic broadcasts that all devices receive in parallel, the concept assumes that all devices are connected to the same infrastructure, i.e., all stations are connected to a single \gls{ap}, using \mbox{Wi-Fi} as communication system. 
In order to achieve a wider service area, \cite{gundall2021extending} proposes a concept for the possible extension of the \gls{rbis} protocol. It is based on the idea that there are \mbox{Wi-Fi} stations that are in the range of more than one \gls{ap} and consequently receiving multiple synchronization broadcasts. If the time offset between these transmissions is considered, also clocks of devices can be synchronized that are only connected to different \glspl{ap}. 

Furthermore, the use of the \gls{rbis} protocol not only in IEEE~802.11 but also mobile radio communications was investigated \cite{gundall2020Integration}. Accordingly, a first draft for using \gls{rbis} protocol in \gls{5g} was proposed, but the concept was evaluated using 4G. Moreover, no performance results on the clock synchronization of the \glspl{ue} were provided. The evaluation was only analysed regarding measurements of a discrete factory automation demonstrator that requires a clock synchronization of 1~ms and less.

\section{Mapping of RBIS to 5G}%
\label{sec:Mapping of RBIS to 5G}
This section covers the application of \gls{rbis} protocol in \gls{5g}. In the beginning, Sec.~\ref{subsec:sync_signal}  describes the provision of a synchronization signal, Sec.~\ref{subsec:runtime_correction} a delay correction for using \gls{rbis} protocol in \gls{5g}, and Sec.~\ref{subsec:implementation} introduces the implemented algorithms.

\subsection{Cyclic Synchronization Signal}
\label{subsec:sync_signal}

In order to map the \gls{rbis} protocol to \gls{5g}, a potential synchronization message has to be identified, whereas this message has to be received by all devices that have to be synchronized. Furthermore, it is important that this message is send cyclically and contains an unique identifier that the next message can be differentiated from the recent one. This is required to assign the timestamp $t_j$ to a specific message. The so-called \gls{pbch} that is located inside the \gls{ssb} alongside \gls{pss} and \gls{sss}, broadcasted in the entire network for initial network access, is well suited for this task. Further, the \gls{ssb} is the only always-on signal in \gls{5g}. This means that each \gls{ue} receives the signal even if it is not connected.

Moreover, the \gls{pbch} contains the so-called \gls{sfn}. The \gls{sfn} is a 10-bit counter variable that is incremented every 10~ms. This means that this value repeats every 1024 ticks and is not unique, however, a repetition of every 10.24~s is viable, if $C_R(t)-C_S(t) \leq 10.24/2 = 5.12$~s.
In addition, the \gls{pbch} is send in an interval that can be adjusted by the \gls{gnb} between 5 and 160~ms, which usually depends on the use of the corresponding carrier. A \gls{ue} expects a new \gls{ssb} at least every 20~ms in 5G NR for cell search in an initial (random) access. 
Thus, the basic requirements necessary for \gls{rbis} with respect to the cyclic synchronization signal are already fulfilled by utilizing existing 5G~NR functionalities.

\subsection{Runtime Correction}
\label{subsec:runtime_correction}

\begin{figure}[htbp]
\centering
  \includegraphics[width=.9\columnwidth]{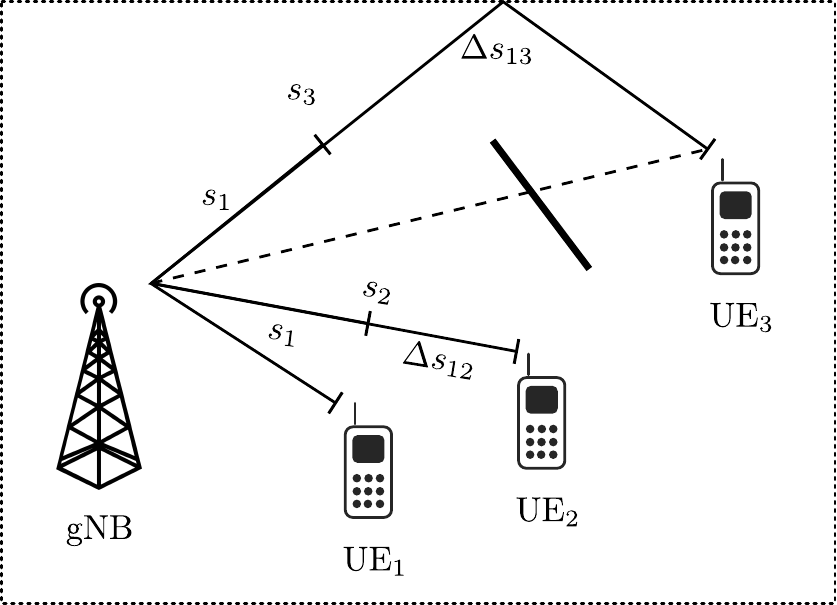} 
  \caption{Time synchronization error due to different \glspl{toa}.}
  \label{fig:error}
\end{figure}

Since the \gls{rbis} protocol is based on the \textit{receiver/receiver} principle, it is assumed that all devices receive the synchronization signal at the same time. This assumption leads to an error depending on the distance deviation $\Delta s$ between \gls{gnb} and the different receivers (see Fig.~\ref{fig:error}).
While $\Delta s$ is small for \glspl{ue} that are equidistant to the \gls{gnb}, a variation of the distance of both \glspl{ue} to the \ac{gnb} leads to an increased $\Delta s$.
Thus, for high values of $\Delta s$, a error correction is required. For mobile communication systems, determining the \gls{ue}-to-\gls{gnb} propagation delay is already of great importance: Due to the provision of frequency- and time-distributed resource elements, the \gls{gnb} expects the packets of the respective \gls{ue} at very strict time slots. So called uplink synchronization must be guaranteed. For smaller time deviations, a cyclic prefix was introduced both between symbols and slots. These are short temporal buffer areas in which no user data is to be received, in order to avoid the overlapping of symbols arriving slightly delayed or ahead of schedule. With large spatial distances and associated relatively long transit times, however, a transmitted OFDM symbol will also exceed the limits of the cyclix prefix by far. Here, the \gls{ta} comes into play \cite{8930825}. \gls{ta} ensures by a time correction that the uplink symbols from the \gls{ue} are sent ahead of time so that they arrive at the \gls{gnb} in the correct time slot. The base station determines the time difference to be set via the uplink signals received and transmits this to the \gls{ue} via the \gls{tac}. The time $T_\mathrm{TA}$ (see Eq.~\ref{eq:timing_advance}), assuming the same signal propagation delay $t_\mathrm{prop}$ in the uplink and downlink, describes in 5G NR the time difference to be considered for sending uplink symbols. In other words: the uplink frame should be sent by the \gls{ue} exactly $T_\mathrm{TA}$ before the corresponding downlink frame from the \gls{gnb} is sent back to the \gls{ue}.

\begin{equation}
\label{eq:timing_advance}
    T_\mathrm{TA}= (N_\mathrm{TA} + N_\mathrm{TA, off}) \cdot T_\mathrm{c} = 2 \cdot t_\mathrm{prop} + T_\mathrm{off} 
\end{equation}

As can be seen in Eq.~\ref{eq:timing_advance}, $T_\mathrm{TA}$ depends not only on the signal propagation delay $t_\mathrm{prop}$ between \gls{ue} and \gls{gnb} but also on an fixed offset $T_\mathrm{off}$. 
The offset is caused by the base station having to switch from receive to transmit between receiving the uplink from and transmitting downlink symbols to the \gls{ue}. 

It should be mentioned that in 5G NR time values are basically specified in a unique step size, the so-called Physical Layer Time Unit $T_\mathrm{c}$. This can be expressed via $T_\mathrm{c} = 1/(\Delta f_\mathrm{max} \cdot N_\mathrm{f}) \approx 0.509~ns$, where $\Delta f_\mathrm{max}$ is the maximum defined \gls{scs} frequency of $480~kHz$ and $N_\mathrm{f}$ the FFT size of 4096. Thus, $T_\mathrm{c}$ can  be considered as 5G NR sampling time. 
The propagation delay $t_\mathrm{prop}$ is also determined in $T_\mathrm{c}$ step size and specified via $N_\mathrm{TA}$. There are two calculation schemes in 5G NR: An initial determination of $N_\mathrm{TA}$ during the random access procedure via the \gls{rar} message as well as continuous updates of $N_\mathrm{TA}$ via the \gls{macce}, which are both specified as \gls{tac} messages. The calculation of $N_\mathrm{TA}$ is performed for the \gls{rar} procedure according to 
\begin{equation}
\label{eq:timing_advance_rar}
    N_\mathrm{TA} = \frac{N_\mathrm{TAC,RAR} \cdot 16 \cdot 64}{2^{\mu}} 
\end{equation}
and for the \gls{macce} updates according to
\begin{equation}
\label{eq:timing_advance_macce}
    N_\mathrm{TA,new} = N_\mathrm{TA, old} + \frac{(N_\mathrm{TAC,MAC}-31) \cdot 16 \cdot 64}{2^{\mu}}.
\end{equation}

The variable $N_\mathrm{TAC}$ is actually the index transmitted as \gls{tac}. $N_\mathrm{TAC}$ in the RAR message has a size of 12 bits and vary between 0 and 3846. In contrast, only a 6 bit $N_\mathrm{TAC}$ index is transmitted in the \gls{macce} that is coded as unsigned and lies between 0 and 63. However, the first 31 values correspond to a 
runtime correction value, as it can be seen by the subtraction in Eq.~\ref{eq:timing_advance_macce}. The parameter $\mu$ in Eq.~\ref{eq:timing_advance_rar}-\ref{eq:timing_advance_macce} indicates the subcarrier spacing used and is based on the following relation: $\Delta f = 2^\mu \cdot 15~kHz$.

Thus, \gls{ta} in 5G~NR operates based on a correction in fixed step sizes.
To get a better overview of correction accuracies, both the temporal and spatial step sizes for different \gls{scs} are shown in Table \ref{tab:timing_advance_step_size}.
\begin{table}[htbp]
    \centering
    \caption{Timing Advance adjustment step size for different subcarrier frequencies}
    \begin{tabular}{c   c   c   c   c}
    \toprule 
    $\mu$ & Freq. & \gls{scs}  & \multicolumn{2}{c}{Adjustment Step Size}\\
     & Range & $\Delta f$ [kHz]& time [ns] & distance [m]\\ 
    \midrule
    0 & \multirow{2}{*}{FR1} & 15  & $520.85$ & $78.13$  \\
    1 &  & 30  & $260.42$  & $39.06$ \\
    \cmidrule{1-5}
    2 & FR1/2 & 60 & $130.21$ & $19.53$ \\
    \cmidrule{1-5}
    3 & FR2 & 120 & $65.10$ & $9.77$\\
      \bottomrule 
    \end{tabular}
    \label{tab:timing_advance_step_size}
\end{table}
The distance in meters is also a metric to figure out which distance an \gls{ue} has to move before the \gls{ta} value changes. The adjustment time step size is also the minimum possible correction time and limits the accuracy that can be achieved in combination with \gls{rbis}. If $\Delta f = 30~kHz$, runtime differences up to twice the corresponding adjustment time can be compensated. In this case, this leads to an accuracy of slightly more than half a microsecond. 

Also for runtime correction, the 5G NR standard with its \gls{ta} functionalities provide a well-founded baseline. For this, however, the master \gls{ue} must exchange its calculated $T_\mathrm{TA}$ value together with its timestamp to the other \glspl{ue}, since the \gls{tac} response packets are only transmitted to the respective \gls{ue} and not broadcasted within the cell. Then, slave \glspl{ue} can perform a RBIS protocol-based synchronization, corrected for the different runtimes, from the combination of their \gls{ta} value, their own timestamp and the received master values. The offset time $T_\mathrm{off}$, on the other hand, is of no significance for the runtime correction, since it can be assumed that this is the same for all transmitted cyclic broadcast signals.


\subsection{Implementation}
\label{subsec:implementation}

In order to implement \gls{rbis} protocol in \gls{5g}, two different algorithms have to be formulated, whereas Alg.~\ref{alg:refue} shows the algorithm for the Master~\gls{ue} $M$. First, it has to be ensured that all Slave~\glspl{ue} $S_x$ 
are connected to the same \gls{gnb}. Furthermore, a reference time $t^R$ is required at $M$. Since the \gls{sfn} repeats every 10.24~s, the clocks of all $S_x$ have to be at least within the already mentioned 5.12~s limit. Thus, an initial timestamp $t_0^R$ is created and send to all $S_x$ alongside the Masters \gls{ta} $T_\mathrm{TA}^R$. Afterwards, $M$ waits for a \gls{pbch} block and creates a timestamp $t^R$, when a new one is detected. Furthermore, both $\gls{sfn}$ and $\mu$ values are read. Then, a new pair $\langle SFN^R,t^R\rangle$ is created and sent to all $S_x$. Since $\mu$ defines the the minimal intervals $T_I$, where all \glspl{ue} receive a new \gls{pbch} block, it does not make sense to send \textit{FOLLOW~UP} messages more frequently. Furthermore, $N_\mathrm{TA}^R$ is included in the \textit{FOLLOW~UP} message.

\begin{algorithm}
\caption{Master UE algorithm}\label{alg:refue}
\begin{algorithmic}[1]
\Require $t^R, x \in\mathbb{N}$, $T_\mathrm{TA}^R$
\Ensure$M$, $S_x$ are in the range of the same BS
\State Create timestamp $t_0^R$ 
\State Send $t_0^R$  to $S_x$, $\forall x$
\While{true} 
\State Wait for PBCH
\State Create timestamp $t^R$
\State Read $\mu$
\State Read $SFN^R$
\State Create pair $\langle SFN^R,t^R\rangle$
\State Transmit $\langle SFN^R,t^R, N_\mathrm{TA}^R\rangle$ to $S_x$, $\forall x$
\EndWhile  
\end{algorithmic}
\end{algorithm}

Thus, the algorithm for the $S_x$ is presented in Alg.~\ref{alg:slaveue}. After its start up, each $S_x$ waits for the initial  timestamp $t_0^R$ of $M$. When it arrives, a timestamp of its local time $t_0^S$ is taken and the difference between both timestamps is calculated. If the timestamp is within the limit of 5.12~s, no adjustment of $C_S(t)$ is required. In the next step, each $S_x$ waits for an incoming \gls{pbch} block and processes it similar to $M$. After the pair  $\langle SFN^S,t^S\rangle$ is created, $S_x$ waits for incoming \textit{FOLLOW~UP} messages. When a \textit{FOLLOW~UP} message is received, the included \gls{sfn} is compared to the most recent one that was included in the last \gls{pbch} block. Dependent on the delay of the processing and transmission of the \textit{FOLLOW~UP} messages by $S_x$, it is possible that they differ. In order to get the suitable pair, $S_x$ browses $\langle SFN_{t-1},t_{j-1}^S\rangle$ pair until it finds the correct one. Afterwards, the offset that is defined as $i$ is calculated and can be used for further iterations. Last but not least, the offset is calculated 
and the clock $C_S(t)$ is corrected.

\begin{algorithm}
\caption{Slave \gls{ue} algorithm}\label{alg:slaveue}
\begin{algorithmic}[1]
\Require $t^S, x \in\mathbb{N}$, $T_\mathrm{TA}^S$
\Ensure$M$, $S_x$ are in the range of the same BS
\State $i \gets 0$
\State Wait for $t_0^R$
\State Create timestamp $t_0^S$
\If{$|t_0^R-t_0^S|\leq5.12$s}
\State $C_S(t) \gets t_0^R $
\EndIf
\While{true} 
\State Wait for PBCH
\State Create timestamp $t^S$
\State Read $\mu$
\State Read $SFN^S$
\State Create pair $\langle SFN^S,t_{SFN}^S\rangle$
\State $SFN^S \gets SFN^S - i$
\State Wait for $\langle SFN^R,t^R,T_\mathrm{TA}^R\rangle $
\If{$SFN^S \neq SFN^R$}
\State $i \gets SFN$
\While{$SFN^S \neq SFN^R$}
\State $i \gets (i-16/2^\mu) \mod{1023}$
\EndWhile
\State $i \gets SFN^S-i$
\State $SFN \gets i$
\EndIf
\State $C_S(t)  \gets C_S(t)+t_{SFN}^R-t_{SFN}^S- (T_\mathrm{TA}^R-T_\mathrm{TA}^S)$
\EndWhile  
\end{algorithmic}
\end{algorithm}

\section{Testbed}%
\label{sec:Testbed}
In order to evaluate the proposed concept a testbed is used that contains the hardware that is listed in Tab.~\ref{tab:hardware}. 

\begin{table}[htbp]
    \centering
    \caption{Hardware configurations}
    \begin{tabular}{l   c   l }
    \toprule
\textbf{\textit{Equipment}} & \textbf{\textit{QTY}} & \textbf{\textit{Specification}}\\
     \midrule
5GC & 1 & Nokia AirFrame server\\
5G BS & 1 & Nokia AirScale platform, AirScale  \\
&&Micro Remote Radio Head \\
5G UE (PC)& 2 & Intel Core i9-9900K, \\
& & 16 GB DDR4, \\
& & 2x Intel i210-AT NICs, \\
& & Ubuntu 20.04.1 LTS 64-bit \\
5G UE (SDR) & 2 & Ettus Research X310 \\
TSN Eval. Kit  & 1 & RAPID-TSNEK-V0001, \\
      & & IEEE~802.1AS\\
      \bottomrule 
    \end{tabular}
    \label{tab:hardware}
\end{table}

Since the used \gls{5gc} is a in-house development for Nokia’s Digital Automation Cloud and operated on a Nokia AirFrame server platform, it is not commercially available. The used \gls{gnb}, on the other side, is commercially distributed by Nokia under the name AirScale. This is important because it directly proves that our concept does not require any adaptation of the functionality of a \gls{gnb}.
Additionally, two Ettus Research X310 \gls{sdr} platforms in combination with two high performance \glspl{pc} are used as \glspl{ue}. Furthermore, each \gls{ue} runs \textit{OpenAirInterface5G} software that is an open source experimentation and prototyping platform compliant with the 
\gls{5g} standard. 
Moreover, a \gls{tsn} evaluation kit is used. It is connected to the Master~\gls{ue} in order to synchronize its clock to the \gls{tsn} time. In order to allow time synchronization, the Master~\gls{ue} runs \textit{LinuxPTP} module that is an open source implementation of IEEE~802.1AS and 802.1AS-REV specifications. To perform offset measurements between the \gls{tsn} time and the synchronized time, the Slave~\gls{ue} is also connected to the \gls{tsn} evaluation kit, for evaluation purposes only.

\section{Evaluation}%
\label{sec:Analysis / Evaluation}

\begin{figure}[tb]
	\centering
		\subfloat[Readings of the measured clock offset between \gls{tsn} Evaluation Kit and Intel NUC mini PC over time.]{\resizebox{\columnwidth}{!}{%
\input{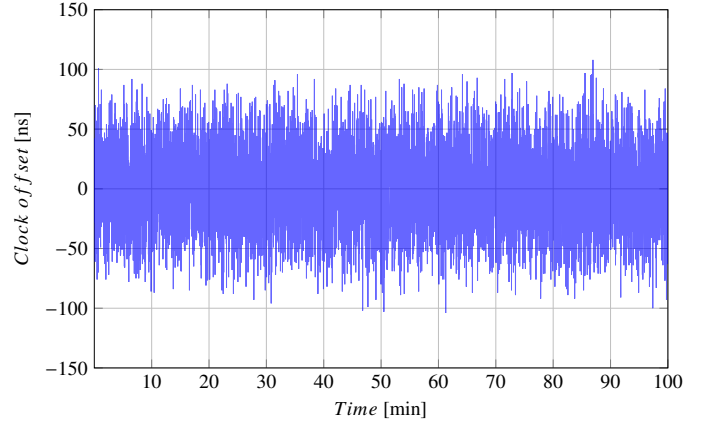}
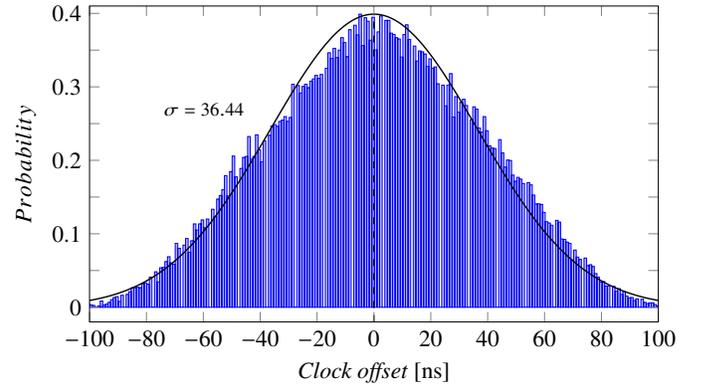}\label{fig:gptp}}

		\subfloat[Histogram of the probability density functions obtained for the readings of the clock offset.]{\resizebox{\columnwidth}{!}{%
\input{figures/gptp_gauss.tikz}}\label{fig:gptpgauss}}
\caption{Clock offset between \gls{tsn} Evaluation Kit and Intel NUC mini PC for a sync interval of 31.25~ms (2\textsuperscript{-5}s).}
\label{fig:gptp_both}
\end{figure}
\begin{figure}[htbp]
\resizebox{\columnwidth}{!}{%
\input{figures/readings_0_5.tikz}}
\caption{Histogram of the reduced data set and probability density functions obtained for different standard deviations and corresponding box plot.}
\label{fig:rbismeasurements}
\end{figure}
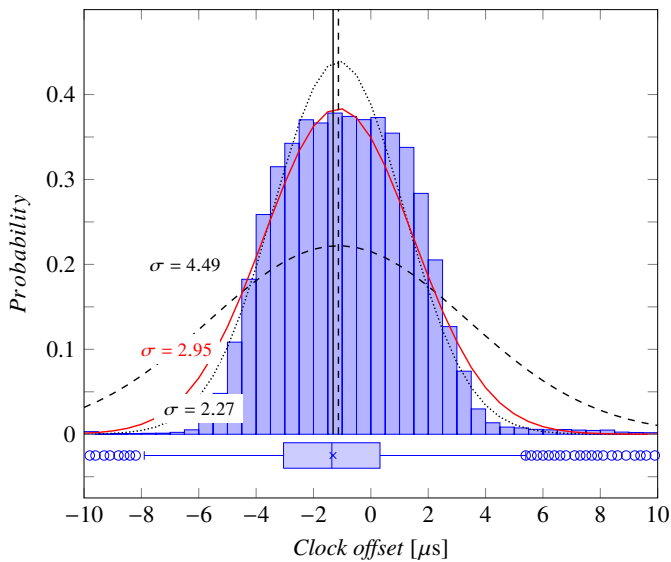

In order to determine the performance of \gls{rbis} protocol using \gls{5g} offset measurements were performed. Since these measurements rely on a very precise measurement of the \gls{tsn} time, Fig.~\ref{fig:gptp_both} depicts offset measurements for the synchronization of the Master~\gls{ue} clock. 
It can be seen, that a maximum synchronization error of $\approx$100~ns occurs, by using the minimum sync interval of 31.25~ms (2\textsuperscript{-5}s). Furthermore, the precision is 0.014$\pm$36.44ns. Sine this value is much lower than the expected clock precision using \gls{rbis} in \gls{5g}, this error can be neglected.
Next, the results of the performed measurements are discussed. Thus, Fig.~\ref{fig:rbismeasurements} shows the histogram of the readings obtained for 35,000 data points. 

Indeed, a total of 180 outliers that exceeded the $\pm$10$~\mu$s borders and are $<$0.5\% of all values are not shown in the plot. Further, three different probability density functions are depicted, obtained for the listed standard deviations. While the dotted density function that belongs to the complete data set is too conservative, the dashed density function for the reduced data set, which contains the remaining 99.5\% of all data points, is too optimistic. Thus, the red curve shows an optimized density function for the measured data set. Using this density function as basis, Tab.~\ref{tab:probabilities} lists the measured clock  precision for different confidence intervals. Furthermore, $P_1$ is the probability value for the reduced data set, while $P_2$ takes the outliers into account.

\begin{table}[htbp]
    \centering
    \caption{Measured clock precision for different standard deviations.}
    \begin{tabular}{c   c   c   c}
    \toprule
   & $\sigma$ & $2\sigma$  & $3\sigma$ \\
     \midrule
  $E(x=\mu)$ [$\mu$s]   & -1.32$\pm$2.95  & -1.32$\pm$5.9 & -1.32$\pm$8.85 \\
  $P_1$ & 0.6827& 0.9545& 0.9973\\
    $P_2$ & 0.6793 & 0.9497 & 0.9937 \\
      \bottomrule 
    \end{tabular}
    \label{tab:probabilities}
\end{table}

\section{Conclusion}%
\label{sec:Conclusion}
In this paper, we highlighted the impact on a precise clock synchronization on upcoming use cases and technologies. Therefore, we proposed the state-of-the-art in wireline as well as wireless communication systems and addressed corresponding challenges. Furthermore, a mapping of the technology independent \gls{rbis} protocol to \gls{5g} was proposed. Moreover, chances and possible drawbacks of the \gls{rbis} protocol were discussed, before a performance evaluation was carried out. The measurements that were performed with real hardware indicate that with only small error corrections, a both precise an accurate wireless clock synchronization ca be performed, taking the very low overhead into account that this protocol requires. Indeed, a clock accuracy of 1.32~$\mu$s with precision of 2.95~$\mu$s was achieved. Thus, approximately 99.5\% of all data values of the offset between master and slave clock lie in the interval of $[-10~\mu s,10~\mu s]$.

Future work on this topic aims to further minimize the clock offset. Here, especially the use of a clock discipline algorithm, e.g. as proposed by \cite{cena2018neural}, improves both accuracy and precision of the clock, by taking not only one but multiple timestamp pairs into account using a filter, or by adopting the frequency of the local oscillators. In this context, the temperature in particular is responsible for frequency variations \cite{6817598}. 


\nobalance
\printbibliography%
\nobalance
\nl
\nobalance
\TempDisplayPreparation
\end{document}

%% file: organization/preamble.tex
\PassOptionsToPackage{usenames,dvipsnames,svgnames,x11names,table,prologue}{xcolor}%
\PassOptionsToPackage{hyphens}{url}%
\PassOptionsToPackage{nomessages}{fp}%
\ifCLASSINFOpdf
  \usepackage[pdftex]{graphicx}
\else
  \usepackage[dvips]{graphicx}
\fi
\ifCLASSOPTIONcompsoc
   \usepackage[caption=false,font=normalsize,labelfont=sf,textfont=sf]{subfig}
\else
   \usepackage[caption=false,font=footnotesize]{subfig}
\fi

\usepackage{stfloats}
\usepackage[english]{babel}%
\usepackage[utf8]{inputenc}%
\usepackage[babel,style=english]{csquotes}%
\usepackage{hyphsubst}%
\usepackage[%
	activate={true,%
	nocompatibility},%
	final,%
	tracking=true,%
	kerning=true,%
	spacing=true,%
	factor=1100,%
	stretch=10,%
	shrink=10%
]{microtype}%
\usepackage{setspace}%
\usepackage{xcolor}%
\definecolor{todonotecol}{RGB}{250,0,0}%
\usepackage{xparse}
\usepackage[%
	colorlinks=false,%
	urlcolor=black,%
	linkcolor=black,%
	citecolor=black,%
	filecolor=black,%
	breaklinks,%
	]{hyperref}%
\usepackage{url}%
\usepackage[%
	acronym,%
	nopostdot,%
	seeautonumberlist,%
	shortcuts,%
	section=chapter,%
	toc,%
]{glossaries}%
\loadglsentries{./supply/glossaries.tex}\glsdisablehyper
\usepackage[%
	backend=biber,%
	style=ieee,%
	isbn=false,%
	hyperref=true,%
	maxbibnames=99,%
	sorting=none,%
	natbib=true,%
	language=english,%
	defernumbers=true,%
	]{biblatex}%
\DeclareFieldFormat{sentencecase}{\csname bbx@colon@search\endcsname#1}

\usepackage{balance}
\usepackage{nameref}
\usepackage{soul}
\usepackage{flushend}
\usepackage{textcomp}
\usepackage{calc}
\usepackage{xkeyval}
\usepackage{multirow}
\usepackage{tabulary}
\usepackage{makecell}
\usepackage{pgfplots}
\usepackage{tikz}
\usepackage{textcomp} 
\usepackage{verbatim}

\pgfplotsset{ grid style={
   line width = 0.1pt
  }, compat=1.16}

%% file: supply/glossaries.tex
\glsaddkey*{url}
{}
{\glsentryurl}
{\Glsentryurl}
{\glsurl}
{\Glsurl}
{\GLSurl}
\glsaddkey*{longpltwo}
{}
{\glsentrylongpltwo}
{\Glsentrylongpltwo}
{\glslongpltwo}
{\Glslongpltwo}
{\GLSlongpltwo}
%
%
%
%
%
\newglossaryentry{}{%
	name={},%
	plural={},%
	description={},%
}%
%
%
%
%
%
%
\newacronym{dmz}{DMZ}{demilitarized zone}
\newacronym{it}{IT}{information technology}
\newacronym{ot}{OT}{operational technology}
\newacronym{opcua}{OPC UA}{Open Platform Communications Unified Architecture}
\newacronym{tsn}{TSN}{time-sensitive networking}
\newacronym{scada}{SCADA}{supervisory control and data acquisition}
\newacronym{vpc}{VPC}{Virtual Process Controller}
\newacronym{plc}{PLC}{programmable logic controller}
\newacronym{udp}{UDP}{User Datagram Protocol}
\newacronym{pubsub}{Pub/Sub}{Publish/Subscribe}
\newacronym{vm}{VM}{virtual machine}
\newacronym{pc}{PC}{personal computer}
\newacronym{ptp}{PTP}{Precision Time Protocol}
\newacronym{ntp}{NTP}{Network Time Protocol}
\newacronym{splc}{Soft-PLC}{Software-PLC}
\newacronym{rte}{RTE}{Real-Time Ethernet}
\newacronym{osi}{OSI}{Open Systems Interconnection}
\newacronym{mac}{MAC}{Media Access Control}
\newacronym{e2e}{E2E}{end-to-end}
\newacronym{tcp}{TCP}{Transmission Control Protocol}
\newacronym{hmi}{HMI}{Human-Machine Interface}
\newacronym{kpi}{KPI}{Key Performance Indicator}
\newacronym{mes}{MES}{Manufacturing Execution System}
\newacronym{rami4.0}{RAMI 4.0}{Reference Architectural Model Industrie 4.0}
\newacronym{tacnet4.0}{TACNET 4.0}{Tactile Internet 4.0}
\newacronym{3gpp}{3GPP}{3rd Generation Partnership Project}
\newacronym{5gppp}{5GPPP}{5G Infrastructure Public Private Partnership}
\newacronym{itu}{ITU}{International Telecommunication Union}
\newacronym{ngnm}{NGNM}{Next Generation Mobile Networks}
\newacronym{agv}{AGV}{automated guided vehicle}
\newacronym{v2v}{V2V}{vehicle-to-vehicle}
\newacronym{v2x}{V2X}{vehicle-to-infrastructure}
\newacronym{d2x}{D2X}{device-to-x}
\newacronym{qos}{QoS}{quality of service}
\newacronym{noa}{NOA}{NAMUR Open Architecture}
\newacronym{dcs}{DCS}{distributed control system}
\newacronym{m2m}{M2M}{machine-to-machine}
\newacronym{sdn}{SDN}{software-defined networking}
\newacronym{erp}{ERP}{enterprise resource planning}
\newacronym{ip}{IP}{Internet Protocol}
\newacronym{lte}{LTE}{Long Term Evolution}
\newacronym{5g}{5G}{5th Generation Wireless Communication Systems} 
\newacronym{harq}{HARQ}{Hybrid Automatic Repeat Request}
\newacronym{rat}{RAT}{radio access technology}
\newacronym{mc}{MC}{Multi-Connectivity}
\newacronym{fbmc}{FBMC}{Filter Bank Multicarrier}
\newacronym{gfdm}{GFDM}{Generalized Frequency Division Multiplexing}
\newacronym{sla}{SLA}{service-level agreement}
\newacronym{5qi}{5QI}{5G QoS Indicator}
\newacronym{bmbf}{BMBF}{German Federal Ministry of Education and Research}
\newacronym{nrt}{NRT}{non real-time}
\newacronym{irt}{IRT}{isochronous real-time}
\newacronym{rt}{RT}{real-time}
\newacronym{ar}{AR}{augmented reality}
\newacronym{wlan}{WLAN}{wireless local area network}
\newacronym{lan}{LAN}{Local Area Network}
\newacronym{cpu}{CPU}{central processing unit}
\newacronym{ram}{RAM}{random-access memory}
\newacronym{ue}{UE}{User Equipment}
\newacronym{ran}{RAN}{Radio Access Network}
\newacronym{bs}{BS}{Base Station}
\newacronym{cn}{CN}{core network}
\newacronym{epc}{EPC}{evolved packet core}
\newacronym{mec}{MEC}{mobile edge cloud}
\newacronym{rtt}{RTT}{round-trip time}
\newacronym{vpn}{VPN}{virtual private network}
\newacronym{vpf}{VPF}{Virtual Process Control Function}
\newacronym{vdi}{VDI}{Association of German Engineers}
\newacronym{ipc}{IPC}{industrial pc}
\newacronym{rtos}{RTOS}{real-time operating system}
\newacronym{vpcmo}{VPC M\&O}{VPC Management \& Orchestration}
\newacronym{i/o}{I/O}{input /output}
\newacronym{fbd}{FBD}{Function Block Diagram}
\newacronym{ldd}{LDD}{Ladder Logic Diagram}
\newacronym{sfc}{SFC}{Sequential Function Chart}
\newacronym{il}{IL}{Instruction List}
\newacronym{st}{ST}{Structured Text}
\newacronym{os}{OS}{operating system}
\newacronym{ir}{IR}{Inactive Resource}
\newacronym{msc}{MSC}{message sequence chart}
\newacronym{macce}{MAC CE}{MAC Control Element}
\newacronym{icps}{ICPS}{industrial cyber-physical system}
\newacronym{cps}{CPS}{cyber-physical systems}
\newacronym{mtd}{MTD}{Moving Target Defense}
\newacronym{iot}{IoT}{Internet of Things}
\newacronym{iiot}{IIoT}{industrial Internet of Things}
\newacronym{iiot2}{IIoT}{Industrial IoT}
\newacronym{k8s}{K8s}{Kubernetes}
\newacronym{rkt}{rkt}{CoreOS Rocket}
\newacronym{dcos}{DC/OS}{Distributed Cloud Operating System}
\newacronym{ml}{ML}{machine learning}
\newacronym{ai}{AI}{artificial intelligence}
\newacronym{ict}{ICT}{industrial communication technology}
\newacronym{kvm}{KVM}{kernel-based virtual machine}
\newacronym{ie}{IE}{Industrial Ethernet}
\newacronym{ias}{IAS}{industrial automation system}
\newacronym{cots}{COTS}{commercial off-the-shelf}
\newacronym{slat}{SLAT}{Second Level Address Translation}
\newacronym{gui}{GUI}{graphical user interface}
\newacronym{l2}{L2}{layer 2}
\newacronym{l3}{L3}{layer 3}
\newacronym{sme}{SME}{small and medium-sized enterprise}
\newacronym{pss}{PSS}{Primary Synchronization Signal}
\newacronym{sss}{SSS}{Secondary Synchronization Signal}
\newacronym{ssb}{SSB}{Synchronization Signal Block}
\newacronym{sfn}{SFN}{System Frame Number}
\newacronym{mib}{MIB}{Master Information Block}
\newacronym{pbch}{PBCH}{Physical Broadcast Channel}
\newacronym{ism}{ISM}{Industrial, Scientific and Medical}
\newacronym{e-utra}{E-UTRA}{evolved UMTS Terrestrial Radio Access}
\newacronym{bldc}{BLDC}{brushless DC}
\newacronym{dl}{DL}{Downlink}
\newacronym{ds-tt}{DS-TT}{Device-Side TSN Translator}
\newacronym{nw-tt}{NW-TT}{Network-Side TSN Translator}
\newacronym{upf}{UPF}{User Plane Function}
\newacronym{tsi}{TSi}{ingress timestamping}
\newacronym{tse}{TSe}{egress timestamping}
\newacronym{gm}{GM}{grandmaster}
\newacronym{gnb}{gNB}{base station}
\newacronym{ta}{TA}{Timing Advance}
\newacronym{tac}{TAC}{Timing Advance Command}
\newacronym{toa}{TOA}{Time of Arrival}
\newacronym{ftsp}{FTSP}{Flooding Time Synchronization Protocol}
\newacronym{rbs}{RBS}{Reference Broadcast Synchronization}
\newacronym{rbis}{RBIS}{Reference Broadcast Infrastructure Synchronization}
\newacronym{rar}{RAR}{Random Access Response}
\newacronym{erbis}{ERBIS}{Extended Reference Broadcast Infrastructure Synchronization}
\newacronym{ap}{AP}{Access Point}
\newacronym{roi}{ROI}{Return On Investment}
\newacronym{scs}{SCS}{Subcarrier Spacing}
\newacronym{los}{LOS}{Line-Of-Sight}
\newacronym{nlos}{NLOS}{Non-Line-Of-Sight}
\newacronym{ru}{RU}{Reference UE}
\newacronym{su}{SU}{Slave UE}
\newacronym{5gc}{5GC}{5G Core}
\newacronym{sdr}{SDR}{Software-Defined Radio}

%% file: organization/makros.tex
\newcommand{\nl}{\par\noindent} 
\newcommand{\mytilde}{{\raise.17ex\hbox{$\scriptstyle\mathtt{\sim}$}}}

%% file: organization/settings.tex
\newlength\textheighttemp%
\newlength\textwidthtemp%
\newlength\textheightstd%
\newlength\textwidthstd%
\newlength\textheightold%
\newlength\textwidthold%
\newlength\tempheight%
\newlength\tempwidth%
\SetProtrusion{encoding={*},family={bch},series={*},size={6,7}}
              {1={ ,750},2={ ,500},3={ ,500},4={ ,500},5={ ,500},
               6={ ,500},7={ ,600},8={ ,500},9={ ,500},0={ ,500}}
\SetExtraKerning[unit=space]
    {encoding={*}, family={bch}, series={*}, size={footnotesize,small,normalsize}}
    {\textendash={400,400}, 
     "28={ ,150}, 
     "29={150, }, 
     \textquotedblleft={ ,150}, 
     \textquotedblright={150, }} 
\SetTracking{encoding={*}, shape=sc}{40}

%% file: organization/IEEE_authors-long.tex
\author{%
\IEEEauthorblockN{%
    Michael Gundall\IEEEauthorrefmark{1}, %
    Julius Stegmann\IEEEauthorrefmark{1}, %
    Christopher Huber\IEEEauthorrefmark{1}, %
    Rüdiger Halfmann\IEEEauthorrefmark{2}, %
    and Hans D. Schotten\IEEEauthorrefmark{1}\IEEEauthorrefmark{3} %
    \\%
}%
\IEEEauthorblockA{%
    \IEEEauthorrefmark{1}German Research Center for Artificial Intelligence GmbH (DFKI), Kaiserslautern, Germany \\%
  \IEEEauthorrefmark{2}Nokia Bell Labs, Munich, Germany \\%
    \IEEEauthorrefmark{3}Department of Electrical and Computer Engineering, Technische Universität Kaiserslautern, Kaiserslautern, Germany %
	\\%
    Email: %
        \{michael.gundall, julius\_raphael.stegmann, christopher.huber,  hans\_dieter.schotten%
        \}@dfki.de
        \\%
      \{ruediger.halfmann\}@nokia-bell-labs.com, %
       \\%
}%
}%


%

%% file: figures/gptp_gauss.tikz
\definecolor{mycolor1}{rgb}{0.00000,0.3,0.6}
\definecolor{mycolor3}{rgb}{0.01,0.79,0.395}

\begin{tikzpicture}
\begin{axis}[ 
width=4in,
height=2.5in,
ymin=-0.02, ymax=.41,
xmin=-100, xmax=100,
area style,
xtick={-100, -80, -60, -40, -20, 0, 20, 40, 60, 80, 100},
ytick={-.5,0, .05, .1,.15,.2,.25,.3,.35,.4},
yticklabels={{},{0},{},{0.1},{},{0.2},{},{0.3},{},{0.4}},
ylabel={$Probability$},
xlabel={\textit{Clock offset} [ns]}
    ]
\addplot+[ybar interval,mark=no] plot coordinates { 
(	-109	,	0.000603544	)
(	-108.0901639	,	0	)
(	-107.1803279	,	0.001207087	)
(	-106.2704918	,	0.001207087	)
(	-105.3606557	,	0	)
(	-104.4508197	,	0.000603544	)
(	-103.5409836	,	0.000603544	)
(	-102.6311475	,	0.001810631	)
(	-101.7213115	,	0.003017718	)
(	-100.8114754	,	0.004828348	)
(	-99.90163934	,	0.003017718	)
(	-98.99180328	,	0.002414174	)
(	-98.08196721	,	0	)
(	-97.17213115	,	0.001810631	)
(	-96.26229508	,	0.00844961	)
(	-95.35245902	,	0.003017718	)
(	-94.44262295	,	0.004224805	)
(	-93.53278689	,	0.006638979	)
(	-92.62295082	,	0.01026024	)
(	-91.71311475	,	0.013277958	)
(	-90.80327869	,	0.014485045	)
(	-89.89344262	,	0.007846066	)
(	-88.98360656	,	0.016295676	)
(	-88.07377049	,	0	)
(	-87.16393443	,	0.017502763	)
(	-86.25409836	,	0.021124024	)
(	-85.3442623	,	0.026555916	)
(	-84.43442623	,	0.026555916	)
(	-83.52459016	,	0.027159459	)
(	-82.6147541	,	0.026555916	)
(	-81.70491803	,	0.031987808	)
(	-80.79508197	,	0.033194895	)
(	-79.8852459	,	0.031384264	)
(	-78.97540984	,	0.041040961	)
(	-78.06557377	,	0	)
(	-77.1557377	,	0.048283483	)
(	-76.24590164	,	0.034401982	)
(	-75.33606557	,	0.053715375	)
(	-74.42622951	,	0.054318919	)
(	-73.51639344	,	0.062164985	)
(	-72.60655738	,	0.068803964	)
(	-71.69672131	,	0.060354354	)
(	-70.78688525	,	0.058543723	)
(	-69.87704918	,	0.08691027	)
(	-68.96721311	,	0.080271291	)
(	-68.05737705	,	0	)
(	-67.14754098	,	0.084496096	)
(	-66.23770492	,	0.093549249	)
(	-65.32786885	,	0.074839399	)
(	-64.41803279	,	0.090531531	)
(	-63.50819672	,	0.113466186	)
(	-62.59836066	,	0.105016576	)
(	-61.68852459	,	0.113466186	)
(	-60.77868852	,	0.118898078	)
(	-59.86885246	,	0.108034294	)
(	-58.95901639	,	0.120105165	)
(	-58.04918033	,	0	)
(	-57.13934426	,	0.133383122	)
(	-56.2295082	,	0.123122882	)
(	-55.31967213	,	0.147264624	)
(	-54.40983607	,	0.152092972	)
(	-53.5	,	0.159939038	)
(	-52.59016393	,	0.179252432	)
(	-51.68032787	,	0.151489429	)
(	-50.7704918	,	0.184684323	)
(	-49.86065574	,	0.205808347	)
(	-48.95081967	,	0.177441801	)
(	-48.04098361	,	0	)
(	-47.13114754	,	0.188909128	)
(	-46.22131148	,	0.203997717	)
(	-45.31147541	,	0.205204804	)
(	-44.40163934	,	0.232364263	)
(	-43.49180328	,	0.201583543	)
(	-42.58196721	,	0.205204804	)
(	-41.67213115	,	0.234778437	)
(	-40.76229508	,	0.214257957	)
(	-39.85245902	,	0.197962281	)
(	-38.94262295	,	0.226328828	)
(	-38.03278689	,	0	)
(	-37.12295082	,	0.234174894	)
(	-36.21311475	,	0.246849308	)
(	-35.30327869	,	0.248659939	)
(	-34.39344262	,	0.246245765	)
(	-33.48360656	,	0.252884744	)
(	-32.57377049	,	0.254695374	)
(	-31.66393443	,	0.259523723	)
(	-30.75409836	,	0.252884744	)
(	-29.8442623	,	0.258316635	)
(	-28.93442623	,	0.303582401	)
(	-28.02459016	,	0	)
(	-27.1147541	,	0.30177177	)
(	-26.20491803	,	0.29151153	)
(	-25.29508197	,	0.300564683	)
(	-24.3852459	,	0.303582401	)
(	-23.47540984	,	0.298754053	)
(	-22.56557377	,	0.307807206	)
(	-21.6557377	,	0.310824923	)
(	-20.74590164	,	0.315653272	)
(	-19.83606557	,	0.311428467	)
(	-18.92622951	,	0.316860359	)
(	-18.01639344	,	0	)
(	-17.10655738	,	0.325309968	)
(	-16.19672131	,	0.346433992	)
(	-15.28688525	,	0.338587926	)
(	-14.37704918	,	0.352469428	)
(	-13.46721311	,	0.339795013	)
(	-12.55737705	,	0.350055254	)
(	-11.64754098	,	0.366954473	)
(	-10.73770492	,	0.360315494	)
(	-9.827868852	,	0.3410021	)
(	-8.918032787	,	0.377818256	)
(	-8.008196721	,	0	)
(	-7.098360656	,	0.383853692	)
(	-6.18852459	,	0.370575734	)
(	-5.278688525	,	0.39894228	)
(	-4.368852459	,	0.356090689	)
(	-3.459016393	,	0.394113932	)
(	-2.549180328	,	0.389285584	)
(	-1.639344262	,	0.363333212	)
(	-0.729508197	,	0.394717476	)
(	0.180327869	,	0.350055254	)
(	1.090163934	,	0.374800539	)
(	2	,	0.396528106	)
(	2.909836066	,	0	)
(	3.819672131	,	0.390492671	)
(	4.729508197	,	0.390492671	)
(	5.639344262	,	0.372989908	)
(	6.549180328	,	0.371179278	)
(	7.459016393	,	0.366954473	)
(	8.368852459	,	0.365143842	)
(	9.278688525	,	0.340398557	)
(	10.18852459	,	0.371179278	)
(	11.09836066	,	0.384457235	)
(	12.00819672	,	0.364540299	)
(	12.91803279	,	0	)
(	13.82786885	,	0.355487145	)
(	14.73770492	,	0.334966665	)
(	15.64754098	,	0.348848167	)
(	16.55737705	,	0.347641079	)
(	17.46721311	,	0.344623362	)
(	18.37704918	,	0.327724143	)
(	19.28688525	,	0.337984383	)
(	20.19672131	,	0.325309968	)
(	21.10655738	,	0.327120599	)
(	22.01639344	,	0.304185945	)
(	22.92622951	,	0	)
(	23.83606557	,	0.302375314	)
(	24.74590164	,	0.274008767	)
(	25.6557377	,	0.302375314	)
(	26.56557377	,	0.318067446	)
(	27.47540984	,	0.258920179	)
(	28.3852459	,	0.286683182	)
(	29.29508197	,	0.265559158	)
(	30.20491803	,	0.278837116	)
(	31.1147541	,	0.283061921	)
(	32.02459016	,	0.274612311	)
(	32.93442623	,	0	)
(	33.8442623	,	0.255298918	)
(	34.75409836	,	0.27280168	)
(	35.66393443	,	0.269783963	)
(	36.57377049	,	0.24383159	)
(	37.48360656	,	0.238399699	)
(	38.39344262	,	0.259523723	)
(	39.30327869	,	0.240210329	)
(	40.21311475	,	0.219689849	)
(	41.12295082	,	0.216672131	)
(	42.03278689	,	0.216068588	)
(	42.94262295	,	0	)
(	43.85245902	,	0.228139458	)
(	44.76229508	,	0.210033152	)
(	45.67213115	,	0.190719759	)
(	46.58196721	,	0.200376456	)
(	47.49180328	,	0.199772912	)
(	48.40163934	,	0.179855975	)
(	49.31147541	,	0.171406366	)
(	50.22131148	,	0.176838257	)
(	51.13114754	,	0.174424083	)
(	52.04098361	,	0.168388648	)
(	52.95081967	,	0	)
(	53.86065574	,	0.169595735	)
(	54.7704918	,	0.166578017	)
(	55.68032787	,	0.152696516	)
(	56.59016393	,	0.140625645	)
(	57.5	,	0.140625645	)
(	58.40983607	,	0.140022101	)
(	59.31967213	,	0.129158318	)
(	60.2295082	,	0.116483903	)
(	61.13934426	,	0.114673273	)
(	62.04918033	,	0.111655555	)
(	62.95901639	,	0	)
(	63.86885246	,	0.118294534	)
(	64.77868852	,	0.11588036	)
(	65.68852459	,	0.092945705	)
(	66.59836066	,	0.092342162	)
(	67.50819672	,	0.0887209	)
(	68.41803279	,	0.08691027	)
(	69.32786885	,	0.064579159	)
(	70.23770492	,	0.077857117	)
(	71.14754098	,	0.06820042	)
(	72.05737705	,	0.065182702	)
(	72.96721311	,	0	)
(	73.87704918	,	0.066993333	)
(	74.78688525	,	0.056733093	)
(	75.69672131	,	0.05794018	)
(	76.60655738	,	0.055526006	)
(	77.51639344	,	0.045869309	)
(	78.42622951	,	0.041040961	)
(	79.33606557	,	0.034401982	)
(	80.24590164	,	0.035005525	)
(	81.1557377	,	0.02897009	)
(	82.06557377	,	0.02897009	)
(	82.97540984	,	0	)
(	83.8852459	,	0.028366546	)
(	84.79508197	,	0.021727567	)
(	85.70491803	,	0.025348829	)
(	86.6147541	,	0.017502763	)
(	87.52459016	,	0.021124024	)
(	88.43442623	,	0.015088589	)
(	89.3442623	,	0.012674414	)
(	90.25409836	,	0.013277958	)
(	91.16393443	,	0.009053153	)
(	92.07377049	,	0.011467327	)
(	92.98360656	,	0	)
(	93.89344262	,	0.010863784	)
(	94.80327869	,	0.006638979	)
(	95.71311475	,	0.004828348	)
(	96.62295082	,	0.006035435	)
(	97.53278689	,	0.006035435	)
(	98.44262295	,	0.003621261	)
(	99.35245902	,	0.001810631	)
(	100.2622951	,	0.001207087	)
(	101.1721311	,	0.002414174	)
(	102.0819672	,	0.001810631	)
(	102.9918033	,	0	)
(	103.9016393	,	0.001207087	)
(	104.8114754	,	0.001810631	)
(	105.7213115	,	0.000603544	)
(	106.6311475	,	0.000603544	)
(	107.5409836	,	0.001207087	)
(	108.4508197	,	0.000603544	)
(	109.3606557	,	0.000603544	)
};

\addplot [color=black, style={semithick}, forget plot]
  table[row sep=crcr]{%
-109	0.004547045	\\
-108.0901639	0.004898122	\\
-107.1803279	0.005273018	\\
-106.2704918	0.00567307	\\
-105.3606557	0.006099671	\\
-104.4508197	0.006554264	\\
-103.5409836	0.007038348	\\
-102.6311475	0.007553475	\\
-101.7213115	0.008101253	\\
-100.8114754	0.008683342	\\
-99.90163934	0.009301454	\\
-98.99180328	0.009957358	\\
-98.08196721	0.010652871	\\
-97.17213115	0.011389864	\\
-96.26229508	0.012170255	\\
-95.35245902	0.012996012	\\
-94.44262295	0.013869149	\\
-93.53278689	0.014791725	\\
-92.62295082	0.01576584	\\
-91.71311475	0.016793635	\\
-90.80327869	0.017877287	\\
-89.89344262	0.019019004	\\
-88.98360656	0.020221029	\\
-88.07377049	0.021485626	\\
-87.16393443	0.022815083	\\
-86.25409836	0.024211706	\\
-85.3442623	0.025677812	\\
-84.43442623	0.027215727	\\
-83.52459016	0.028827777	\\
-82.6147541	0.030516285	\\
-81.70491803	0.032283563	\\
-80.79508197	0.034131906	\\
-79.8852459	0.036063587	\\
-78.97540984	0.038080846	\\
-78.06557377	0.040185886	\\
-77.1557377	0.042380864	\\
-76.24590164	0.04466788	\\
-75.33606557	0.047048976	\\
-74.42622951	0.04952612	\\
-73.51639344	0.052101199	\\
-72.60655738	0.054776014	\\
-71.69672131	0.057552265	\\
-70.78688525	0.060431547	\\
-69.87704918	0.063415335	\\
-68.96721311	0.066504979	\\
-68.05737705	0.069701693	\\
-67.14754098	0.073006542	\\
-66.23770492	0.076420438	\\
-65.32786885	0.079944126	\\
-64.41803279	0.083578175	\\
-63.50819672	0.087322971	\\
-62.59836066	0.091178703	\\
-61.68852459	0.095145359	\\
-60.77868852	0.099222714	\\
-59.86885246	0.10341032	\\
-58.95901639	0.107707502	\\
-58.04918033	0.112113347	\\
-57.13934426	0.116626697	\\
-56.2295082	0.12124614	\\
-55.31967213	0.125970009	\\
-54.40983607	0.130796369	\\
-53.5	0.135723018	\\
-52.59016393	0.140747477	\\
-51.68032787	0.14586699	\\
-50.7704918	0.151078517	\\
-49.86065574	0.156378736	\\
-48.95081967	0.161764036	\\
-48.04098361	0.16723052	\\
-47.13114754	0.172774003	\\
-46.22131148	0.178390016	\\
-45.31147541	0.184073803	\\
-44.40163934	0.189820326	\\
-43.49180328	0.195624272	\\
-42.58196721	0.20148005	\\
-41.67213115	0.207381807	\\
-40.76229508	0.213323426	\\
-39.85245902	0.219298538	\\
-38.94262295	0.22530053	\\
-38.03278689	0.231322556	\\
-37.12295082	0.237357545	\\
-36.21311475	0.243398217	\\
-35.30327869	0.249437092	\\
-34.39344262	0.255466505	\\
-33.48360656	0.261478624	\\
-32.57377049	0.26746546	\\
-31.66393443	0.273418888	\\
-30.75409836	0.279330662	\\
-29.8442623	0.285192433	\\
-28.93442623	0.290995771	\\
-28.02459016	0.296732181	\\
-27.1147541	0.302393123	\\
-26.20491803	0.307970036	\\
-25.29508197	0.313454354	\\
-24.3852459	0.318837535	\\
-23.47540984	0.324111073	\\
-22.56557377	0.329266529	\\
-21.6557377	0.334295549	\\
-20.74590164	0.339189886	\\
-19.83606557	0.343941423	\\
-18.92622951	0.348542198	\\
-18.01639344	0.352984421	\\
-17.10655738	0.357260501	\\
-16.19672131	0.361363062	\\
-15.28688525	0.365284971	\\
-14.37704918	0.369019351	\\
-13.46721311	0.372559609	\\
-12.55737705	0.375899449	\\
-11.64754098	0.379032892	\\
-10.73770492	0.381954298	\\
-9.827868852	0.384658376	\\
-8.918032787	0.387140207	\\
-8.008196721	0.389395253	\\
-7.098360656	0.391419374	\\
-6.18852459	0.393208842	\\
-5.278688525	0.394760349	\\
-4.368852459	0.396071017	\\
-3.459016393	0.397138411	\\
-2.549180328	0.397960543	\\
-1.639344262	0.398535881	\\
-0.729508197	0.398863349	\\
0.180327869	0.398942336	\\
1.090163934	0.398772694	\\
2	0.39835474	\\
2.909836066	0.397689255	\\
3.819672131	0.396777482	\\
4.729508197	0.395621119	\\
5.639344262	0.394222319	\\
6.549180328	0.39258368	\\
7.459016393	0.390708235	\\
8.368852459	0.388599448	\\
9.278688525	0.3862612	\\
10.18852459	0.383697777	\\
11.09836066	0.380913856	\\
12.00819672	0.377914495	\\
12.91803279	0.374705114	\\
13.82786885	0.371291478	\\
14.73770492	0.367679685	\\
15.64754098	0.363876139	\\
16.55737705	0.359887542	\\
17.46721311	0.355720864	\\
18.37704918	0.351383331	\\
19.28688525	0.346882399	\\
20.19672131	0.342225733	\\
21.10655738	0.33742119	\\
22.01639344	0.332476791	\\
22.92622951	0.327400702	\\
23.83606557	0.322201212	\\
24.74590164	0.31688671	\\
25.6557377	0.31146566	\\
26.56557377	0.305946583	\\
27.47540984	0.300338035	\\
28.3852459	0.294648582	\\
29.29508197	0.288886778	\\
30.20491803	0.28306115	\\
31.1147541	0.277180171	\\
32.02459016	0.271252245	\\
32.93442623	0.265285686	\\
33.8442623	0.259288695	\\
34.75409836	0.253269352	\\
35.66393443	0.247235589	\\
36.57377049	0.241195181	\\
37.48360656	0.235155725	\\
38.39344262	0.22912463	\\
39.30327869	0.223109103	\\
40.21311475	0.217116132	\\
41.12295082	0.211152481	\\
42.03278689	0.205224673	\\
42.94262295	0.199338988	\\
43.85245902	0.193501447	\\
44.76229508	0.187717809	\\
45.67213115	0.181993563	\\
46.58196721	0.176333922	\\
47.49180328	0.170743823	\\
48.40163934	0.165227915	\\
49.31147541	0.159790567	\\
50.22131148	0.154435857	\\
51.13114754	0.149167579	\\
52.04098361	0.143989236	\\
52.95081967	0.13890405	\\
53.86065574	0.133914954	\\
54.7704918	0.129024605	\\
55.68032787	0.124235379	\\
56.59016393	0.119549382	\\
57.5	0.114968448	\\
58.40983607	0.110494152	\\
59.31967213	0.106127812	\\
60.2295082	0.101870495	\\
61.13934426	0.097723028	\\
62.04918033	0.093686002	\\
62.95901639	0.089759782	\\
63.86885246	0.085944514	\\
64.77868852	0.082240137	\\
65.68852459	0.078646387	\\
66.59836066	0.075162813	\\
67.50819672	0.071788778	\\
68.41803279	0.068523476	\\
69.32786885	0.065365939	\\
70.23770492	0.062315045	\\
71.14754098	0.059369531	\\
72.05737705	0.056527998	\\
72.96721311	0.053788928	\\
73.87704918	0.051150686	\\
74.78688525	0.048611535	\\
75.69672131	0.04616964	\\
76.60655738	0.043823084	\\
77.51639344	0.041569871	\\
78.42622951	0.039407937	\\
79.33606557	0.037335161	\\
80.24590164	0.035349367	\\
81.1557377	0.033448338	\\
82.06557377	0.031629821	\\
82.97540984	0.029891535	\\
83.8852459	0.028231178	\\
84.79508197	0.026646432	\\
85.70491803	0.025134973	\\
86.6147541	0.023694474	\\
87.52459016	0.022322612	\\
88.43442623	0.021017074	\\
89.3442623	0.019775559	\\
90.25409836	0.018595788	\\
91.16393443	0.017475503	\\
92.07377049	0.016412475	\\
92.98360656	0.015404506	\\
93.89344262	0.014449431	\\
94.80327869	0.013545125	\\
95.71311475	0.012689502	\\
96.62295082	0.011880519	\\
97.53278689	0.01111618	\\
98.44262295	0.010394534	\\
99.35245902	0.009713679	\\
100.2622951	0.009071764	\\
101.1721311	0.00846699	\\
102.0819672	0.007897609	\\
102.9918033	0.007361928	\\
103.9016393	0.006858304	\\
104.8114754	0.006385151	\\
105.7213115	0.005940937	\\
106.6311475	0.005524182	\\
107.5409836	0.005133462	\\
108.4508197	0.004767404	\\
109.3606557	0.004424691	\\
110.2704918	0.004104055	\\
111.1803279	0.003804282	\\
112.0901639	0.003524207	\\
};
\addplot [color=black, dashed, style={semithick},  forget plot]
  table[row sep=crcr]{%
  -0.014    -0.02 \\
  -0.014    0.40\\
 };
\node[at={(-60,.27)},fill= white, text=black] {\footnotesize{$\sigma=36.44$}};
\end{axis}
\end{tikzpicture}

%% file: figures/readings_0_5.tikz
\definecolor{mycolor1}{rgb}{0.00000,0.3,0.6}
\definecolor{mycolor3}{rgb}{0.01,0.79,0.395}

\begin{tikzpicture}
\begin{axis}[
width=4in,
height=3.5in,
ymin=-0.015, ymax=.1,
xmin=-10, xmax=10,
area style,
xtick={-10, -8, -6, -4, -2, 0, 2, 4, 6, 8, 10},
ytick={-.01,0, .01, .02,.03,.04,.05,.06,.07,.08},
yticklabels={{},{0},{},{0.1},{},{0.2},{},{0.3},{},{0.4}},
ylabel={$Probability$},
xlabel={\textit{Clock offset} [$\mu$s]}
    ]
\addplot+[ybar interval,mark=no] plot coordinates { 
(	-10	,	0.000619038	)
(	-9.5	,	5.63E-05	)
(	-9	,	0.000140691	)
(	-8.5	,	0.000168829	)
(	-8	,	0.000225105	)
(	-7.5	,	0.000253243	)
(	-7	,	0.00045021	)
(	-6.5	,	0.00104111	)
(	-6	,	0.00340471	)
(	-5.5	,	0.009623231	)
(	-5	,	0.021694476	)
(	-4.5	,	0.036495118	)
(	-4	,	0.051745969	)
(	-3.5	,	0.06300121	)
(	-3	,	0.068516278	)
(	-2.5	,	0.074059484	)
(	-2	,	0.073299755	)
(	-1.5	,	0.075635218	)
(	-1	,	0.074847351	)
(	-0.5	,	0.074087622	)
(	0	,	0.074594108	)
(	0.5	,	0.070908017	)
(	1	,	0.067559582	)
(	1.5	,	0.056726413	)
(	2	,	0.041053491	)
(	2.5	,	0.02535243	)
(	3	,	0.014856918	)
(	3.5	,	0.005965278	)
(	4	,	0.002701258	)
(	4.5	,	0.001688286	)
(	5	,	0.001491319	)
(	5.5	,	0.001153662	)
(	6	,	0.00104111	)
(	6.5	,	0.001125524	)
(	7	,	0.000984834	)
(	7.5	,	0.000900419	)
(	8	,	0.000984834	)
(	8.5	,	0.000506486	)
(	9	,	0.000422072	)
(	9.5	,	0.000365795	)
(	10	,	0.000253243	)
};

\addplot [color=black, style={semithick}, dashed, forget plot]
  table[row sep=crcr]{%
-10		0.006313128	\\
-9.5		0.007816782	\\
-9		0.009559421	\\
-8.5		0.011546632	\\
-8		0.013775241	\\
-7.5		0.016231671	\\
-7		0.018890673	\\
-6.5		0.021714598	\\
-6		0.024653371	\\
-5.5		0.027645278	\\
-5		0.030618632	\\
-4.5		0.033494289	\\
-4		0.036188942	\\
-3.5		0.038619013	\\
-3		0.040704892	\\
-2.5		0.042375243	\\
-2		0.043571042	\\
-1.5		0.04424904	\\
-1		0.044384355	\\
-0.5		0.04397199	\\
0		0.043027141	\\
0.5		0.041584264	\\
1		0.03969499	\\
1.5		0.037425062	\\
2		0.034850541	\\
2.5		0.032053589	\\
3		0.029118162	\\
3.5		0.026125909	\\
4		0.02315256	\\
4.5		0.020265008	\\
5		0.017519218	\\
5.5		0.014959007	\\
6		0.01261569	\\
6.5		0.010508468	\\
7		0.008645457	\\
7.5		0.007025167	\\
8		0.005638265	\\
8.5		0.004469454	\\
9		0.00349932	\\
9.5		0.002706032	\\
10		0.002066819	\\
};

\addplot [color=black,style={semithick},  densely dotted,forget plot]
  table[row sep=crcr]{%
-10		4.23E-05	\\
-9.5		9.76E-05	\\
-9		0.000214602	\\
-8.5		0.000449658	\\
-8		0.000897556	\\
-7.5		0.001706751	\\
-7		0.003091779	\\
-6.5		0.005335515	\\
-6		0.0087715	\\
-5.5		0.013737286	\\
-5		0.020495459	\\
-4.5		0.029130233	\\
-4		0.039442085	\\
-3.5		0.050875113	\\
-3		0.062514465	\\
-2.5		0.073178796	\\
-2		0.081605533	\\
-1.5		0.086692906	\\
-1		0.087735857	\\
-0.5		0.084586352	\\
0		0.077687846	\\
0.5		0.067972848	\\
1		0.056656203	\\
1.5		0.044987211	\\
2		0.03402987	\\
2.5		0.024522295	\\
3		0.016834164	\\
3.5		0.011009094	\\
4		0.006858691	\\
4.5		0.004070619	\\
5		0.002301491	\\
5.5		0.001239618	\\
6		0.000636057	\\
6.5		0.000310909	\\
7		0.000144777	\\
7.5		6.42E-05	\\
8		2.71E-05	\\
8.5		1.09E-05	\\
9		4.19E-06	\\
9.5		1.53E-06	\\
10		5.33E-07	\\
  };
  
\addplot [color=red, style={semithick},  forget plot]
  table[row sep=crcr]{%
-10		0.000226866	\\
-9.5		0.000429308	\\
-9		0.0007829	\\
-8.5		0.001375888	\\
-8		0.002330229	\\
-7.5		0.003803232	\\
-7		0.005981993	\\
-6.5		0.009067297	\\
-6		0.013244897	\\
-5.5		0.018644817	\\
-5		0.025293357	\\
-4.5		0.033066912	\\
-4		0.041660034	\\
-3.5		0.050580652	\\
-3		0.059181781	\\
-2.5		0.066731433	\\
-2		0.072512301	\\
-1.5		0.075933206	\\
-1		0.076628549	\\
-0.5		0.074522648	\\
0		0.069843303	\\
0.5		0.063081219	\\
1		0.05490529	\\
1.5		0.046053975	\\
2		0.037227071	\\
2.5		0.028999428	\\
3		0.021770019	\\
3.5		0.015749508	\\
4		0.010980294	\\
4.5		0.00737734	\\
5		0.004776661	\\
5.5		0.002980491	\\
6		0.001792215	\\
6.5		0.001038559	\\
7		0.000579977	\\
7.5		0.000312126	\\
8		0.000161878	\\
8.5		8.09E-05	\\
9		3.90E-05	\\
9.5		1.81E-05	\\
10		8.09E-06	\\
};  

\addplot [color=black, style={semithick},  forget plot]
  table[row sep=crcr]{%
  -1.316    0.0 \\
  -1.316    0.40\\
 };
 
 \addplot [color=black,dashed, style={semithick}, forget plot]
  table[row sep=crcr]{%
  -1.13   0.0 \\
  -1.13    0.40 \\
 };
 
 \node[at={(-6.5,.04)},fill= white, text=black] {\footnotesize{$\sigma=4.49$}};
 \node[at={(-6,.006)}, fill= white, text=black] {\footnotesize{$\sigma=2.27$}};
 \node[at={(-6.8,.02)},fill= white, text=red] {\footnotesize{$\sigma=2.95$}};
\addplot [color=blue, forget plot]
  table[row sep=crcr]{%
5.4   -0.004\\
5.4   -0.006\\
};
\addplot [color=blue, forget plot]
  table[row sep=crcr]{%
-7.9 -0.004\\
-7.9 -0.006\\
};
\addplot [color=blue, forget plot]
  table[row sep=crcr]{%
5.4 -0.005\\
0.319 -0.005\\
};
\addplot [color=blue, forget plot]
  table[row sep=crcr]{%
-7.9 -0.005\\
-3.044 -0.005\\
};
\addplot [color=blue, ,fill=blue, fill opacity=0.2, forget plot]
  table[row sep=crcr]{%
-3.044 -0.008\\
0.319   -0.008\\
0.319   -0.002\\
-3.044 -0.002\\
-3.044 -0.008\\
};
\addplot [color=blue, forget plot]
  table[row sep=crcr]{%
-1.36	-0.008\\
-1.36	-0.002\\
};
\addplot [color=blue, only marks, mark=x, mark options={solid, draw=blue}, forget plot]
  table[row sep=crcr]{%
-1.32	-0.005\\
};
\addplot [color=blue, only marks, mark=o, mark options={solid, draw=blue}, forget plot]
  table[row sep=crcr]{%
-9.8	-0.005\\
-9.6	-0.005\\
-9.3	-0.005\\
-9.1	-0.005\\
-8.8	-0.005\\
-8.6	-0.005\\
-8.2	-0.005\\
-8.4	-0.005\\
5.4	-0.005\\
5.6	-0.005\\
5.8	-0.005\\
6.0	-0.005\\
6.2	-0.005\\
6.4	-0.005\\
6.6	-0.005\\
6.8	-0.005\\
7.1	-0.005\\
7.3	-0.005\\
7.5	-0.005\\
7.7	-0.005\\
7.9	-0.005\\
8.1	-0.005\\
8.4	-0.005\\
8.6	-0.005\\
8.9	-0.005\\
9.2	-0.005\\
9.4	-0.005\\
9.6	-0.005\\
9.9	-0.005\\
  };
\end{axis}
\end{tikzpicture}